\documentclass[runningheads, a4paper]{llncs}
    \usepackage{lineno}
	\usepackage{fancyvrb}
	\usepackage{color}

	\usepackage{graphicx}
	\usepackage{amssymb, amsmath}
	\usepackage{enumerate}
	\usepackage[noend]{algpseudocode}
	\usepackage{algorithm}
	 \usepackage[colorlinks]{hyperref}
	\usepackage{xspace}

	\algdef{SE}[DOWHILE]{Do}{doWhile}{\algorithmicdo}[1]{\algorithmicwhile\ #1}
	
	\newcommand {\R}   {\mathbb R}
	
	\newcommand {\C}   {\mathbb C}
	
	\newcommand {\N}   {\mathbb N}
	
	\newcommand {\ii}  {{\mathbf i}}
	\newcommand{\conj}[1]{\overline{#1}}
	\newcommand{\conjclo}[1]{#1_{\conj{\cup}}}
	\newcommand{\mIm}[1]{Im(#1)}
	
    \newcommand{\solsIn}[2]{\texttt{Zero}(#1)}
    \newcommand{\nbSolsIn}[2]{\#(#1)}
	\newcommand{\width}[1]{w(#1)}
	\newcommand{\contDisc}[1]{\Delta(#1)}

	\newcommand{\ccomp}[1]{\mathcal{#1}}
	\newcommand{\compBox}[1]{B_{#1}}
	
	\newcommand{\ass}{\leftarrow}
	\newcommand{\app}[1]{\widetilde{#1}}

	\newcommand{\Tstart}{T^*}
	\newcommand{\Czerot}{C^0}
	\newcommand{\Cstart}{C^*}
	\newcommand{\Pstart}{P^*}
	\newcommand{\PstartApp}{\app{P^*}}
	\newcommand{\Tstar}[1]{T^*(#1)}
	\newcommand{\Czero}[1]{C^0(#1)}
	\newcommand{\Cstar}[1]{C^*(#1)}
	\newcommand{\Pstar}[1]{P^*(#1)}
	\newcommand{\PstarApp}[1]{\app{P^*}(#1)}

	\newcommand{\calO}{{\mathcal{O}}}
	\newcommand{\calI}{{\mathcal{I}}}
	\newcommand{\intbox}{%
	  {\,\setlength{\unitlength}{.33mm}\framebox(4,7){}\,}}

	\newcommand{\RI}[1]{{\color{magenta}#1}}
	\renewcommand{\RI}[1]{{#1}}
	\newcommand{\RIn}[1]{{\color{blue}#1}}
	\renewcommand{\RIn}[1]{{#1}}

	\newcommand{\coblue}[1]{{\color{blue}#1}}
	
	\newcommand{\ignore}[1]{}
	
	\newcommand{\ccluster}{\texttt{Ccluster}\xspace}
	\newcommand{\cclusterO}{\texttt{CclusterO}\xspace}
	\newcommand{\cclusterR}{\texttt{CclusterR}\xspace}
	
	\newcommand{\cclusterPs}{\texttt{CclusterPs}\xspace}
	
	\newcommand{\clang}{\texttt{C}\xspace}
	\newcommand{\mpsolve}{\texttt{MPsolve}\xspace}
	
	\newcommand{\Ber}[1]{\mbox{\tt Bernoulli}_{#1}}
	\newcommand{\Mig}[2]{\mbox{\tt Mignotte}_{#2}}
	\newcommand{\Man}[1]{\mbox{\tt Mandelbrot}_{#1}}
	\newcommand{\Run}[1]{\mbox{\tt Runnels}_{#1}}

\usepackage{breqn}
\newtheorem{Definition}{Definition}%[section]
\newtheorem{Remark}[Definition]{Remark}
\newtheorem{Theorem}[Definition]{Theorem}

\newtheorem{Proposition}[Definition]{Proposition}

 \title{
New  practical advances in polynomial root clustering}
 \titlerunning{Some practical advances in polynomial root clustering}
\author{R\'emi Imbach \inst{1}
		\thanks{
		R\'emi's work is supported by NSF Grants \#~CCF-1563942
    and \#~CCF-1708884.
		}
	\and Victor Y. Pan\inst{2}
		\thanks{
	    Victor's work is supported by NSF Grants
	    \#~CCF-1116736 and \#~CCF-1563942 and by PSC CUNY Award
		698130048.}
	    }
\authorrunning{Imbach-Pan}
\institute{
  Courant Institute of Mathematical Sciences\\
  New York University, USA\\
  Email: \email{remi.imbach@nyu.edu}\\
  \url{https://cims.nyu.edu/~imbach/}
\and
  City University of New York, USA\\
  Email: \email{victor.pan@lehman.cuny.edu}\\
  \url{http://comet.lehman.cuny.edu/vpan/}
}

\graphicspath{{./figures/png/}} 

\begin{document}
%%%%%%%%%%%%%%%%%%%%%%%%%%%%%%%%%%%%%%%%%%%%%%
\maketitle
 
\begin{abstract}
We report an ongoing work on clustering
    algorithms for complex roots of a univariate polynomial 
      $p$ of degree $d$  with real or complex coefficients. As in their previous  best 
      subdivision algorithms  our root-finders
      are robust 
      even for  multiple roots
      of a polynomial given by a black box for the
      approximation of its coefficients, and 
      their complexity decreases at least proportionally to the number of roots in a region of interest (ROI)
       on the complex plane, such as a disc or a square,
      but we greatly strengthen the main ingredient of the previous algorithms. Namely our
      new counting test  essentially amounts to the evaluation of a polynomial  $p$ and its derivative $p'$,
 which is a major benefit, e.g., for sparse
      polynomials $p$. 
      Moreover with evaluation at about $\log(d)$
             points (versus the previous record 
of order $d$) we output correct number of 
      roots in a disc whose contour has no roots of $p$ nearby.  Moreover we greatly soften the latter requirement versus the known subdivision algorithms. 
      Our second and less significant contribution concerns
      subdivision algorithms for polynomials with real coefficients. Our tests demonstrate the  power
      of the proposed algorithms.
      
\end{abstract}
% \linenumbers
%%%%%%%%%%%%%%%%%%%%%%%%%%%%%%%%%%%%%%%%%%%%%%
\section{Introduction}
%%%%%%%%%%%%%%%%%%%%%%%%%%%%%%%%%%%%%%%%%%%%%%

We seek complex roots 
of a degree $d$ univariate polynomial $p$  
with real or complex coefficients. For a while the user choice for this problem has been (\RIn{the package \mpsolve}) based  on \emph{e.g.} Erhlich-Aberth
(simultaneous Newton-like) iterations.
Their empirical global convergence
(right from the start) is very fast, but its formal support is a long-known challenge, and 
the iterations approximate 
the roots in a fixed region of interest (ROI) 
about as slow as all complex roots.

In contrast, for the known algorithms subdividing 
 a ROI, e.g., box,  
the cost of root-finding in a ROI decreases
 at least proportionally to the  
 number of roots in it. Some
recent subdivision algorithms have a proved nearly optimal complexity,
are robust 
% even 
in the case of
root clusters and multiple roots, and
their implementation in \cite{ICMSpaper}
 a little outperforms \mpsolve 
for ROI containing only a small number of roots,
which is an important benefit  in many computational areas.

\vspace*{-1em}
\subsubsection*{The Local Clustering Problem}

For a complex set $\mathcal{S}$, 
$\solsIn{\mathcal{S},p}{p}$, or sometimes
 $\solsIn{\mathcal{S}}{p}$, stands for 
 the roots of $p$ in $\mathcal{S}$.
$\nbSolsIn{\mathcal{S},p}{p}$ (or $\nbSolsIn{\mathcal{S}}{p}$) 
stands for the number of roots of $p$ in $\mathcal{S}$.
Here and hereafter the roots are
counted with their multiplicity.

We consider boxes (that is, squares with horizontal and 
vertical edges, parallel to coordinate axis) and
discs 
\RIn{$D(c,r)=\{z\text{ s.t. }|z-c|\leq r\}$}
on the complex plane.
For  such a box (resp. disc) $\mathcal{S}$
and a positive  $\delta$
we denote by $\delta\mathcal{S}$ its concentric
$\delta$-dilation.
We call a disc $\Delta$  an \emph{isolator}
if $\nbSolsIn{\Delta}{p}>0$ and call it 
 \emph{natural} isolator if in addition
$\nbSolsIn{\Delta}{p}=\nbSolsIn{3\Delta}{p}$.
A set $\mathcal{R}$ of roots of $p$ is called a \emph{natural cluster} if there exists
a natural isolator $\Delta$ with
$\solsIn{\mathcal{R}}{p}=\solsIn{\Delta}{p}$.
The Local Clustering Problem (LCP) is to compute natural isolators for natural
clusters together with the sum of multiplicities of roots in the
clusters:

\begin{center} \fbox{
		\begin{minipage}{0.9 \textwidth} \noindent
		\textbf{Local Clustering Problem (LCP):}\\ \noindent
		\textbf{Given:} a polynomial $p\in\C[z]$, 
		%VP
		                a 
% 		                complex square box 
		                ROI
		                $B_0\subset\C$, 
		                $\epsilon>0$\\
		\noindent \textbf{Output:} a set of pairs
		$\{(\Delta^1,m^1),\ldots,(\Delta^\ell,m^\ell)\}$ where: \\ \noindent
		\hphantom{\textbf{Output:}} - the $\Delta^j$'s are pairwise disjoint
		discs of radius $\leq\epsilon$,\\ \noindent
		\hphantom{\textbf{Output:}} -  $m^j =
		\nbSolsIn{\Delta^j,p}{p}=\nbSolsIn{3\Delta^j,p}{p}$ 
		and $m^j>0$ for $j=1,\dots,\ell$
		%VP
		\\ \noindent
		\hphantom{\textbf{Output:}} - $\solsIn{B_0,p}{p} \subseteq
		\bigcup_{j=1}^{\ell} \solsIn{\Delta^j,p}{p} \subseteq
		\solsIn{2B_0,p}{p}$.
	\end{minipage}
	} \end{center}
	
 The basic tool of the nearly optimal subdivision algorithm of \cite{2016Becker}
for  the LCP is the
$\Tstart$-{\em test} 
 for counting the roots of $p$ 
in a complex disc (with multiplicity).
It relies on  Pellet's theorem,
involves approximations
of the coefficients of $p$, and applies
shifting and scaling 
the variable $z$ and 
Dandelin-Gr{\"a}ffe's
root-squaring  iterations.
\cite{ICMSpaper} describes high-level 
improvement
% {\bf REMI, do you mean improvement or implementation?} 
of this test,
and \ccluster\footnote{\url{https://github.com/rimbach/Ccluster}}, 
a \clang implementation of \cite{2016Becker}.

\vspace*{-1em}
\subsubsection*{Our contributions} 
%We propose two contributions to the LCP.
%First, w
Our new counting test, the $\Pstart$-{\em test},
  for a pair of complex $c$ and positive $r$
computes the number $s_0$ of roots of $p$ 
in a complex disc $\Delta$ centered at $c$ with radius $r$. 
If the boundary $\partial\Delta$ contains no roots of $p$, then

\begin{equation}
\label{eq:cauchy}
 s_0 = \frac{1}{2\pi \ii} \int_{\partial\Delta} \frac{p'(z)}{p(z)}dz,~{\rm for}~\ii=\sqrt{-1},
\end{equation}
by virtue of Cauchy's theorem. 
By following \cite{schonhage1982fundamental} and \cite{pan2018old},
 we approximate $s_0$ by
$s_0^*$ obtained by evaluating $p'/p$ on $q$ points on the boundary $\partial\Delta$ within
 the error bound $|s_0 - s_0^*|$ in terms of $q$ 
and  the relative width of a root-free annulus around $\partial\Delta$.
Namely if $\nbSolsIn{\frac{1}{2}\Delta}{p}=\nbSolsIn{2\Delta}{p}$
then for 
$q=\lceil\log_2(d+4) + 2\rceil$
% \RIn{$q< \log_2(2\deg(p)+1)+1$}
% \RIc{check this; my I think $q\simeq \log_2(\deg(p))$}
we recover 
exact value of $s_0$
from $s_0^*$. 

We give an \emph{effective}\footnote{by 
effective, we refer to the pathway proposed in \cite{xu2019effective}
to describe algorithms in three levels: abstract, interval, effective} 
(\emph{i.e.} implementable) 
description of this $\Pstart$-test, which 
 involves
no coefficients of $p$ and can be applied to a polynomial $p$ represented by a black box for its evaluation.  
For sparse 
polynomials and polynomials defined by recursive process such as
Mandelbrot's polynomials (see \cite{bini2000design}, or Eq.~(\ref{eq:mandel}) below),
the test is
particularly efficient and
 the resulting acceleration 
of the clustering algorithm of \cite{2016Becker}
is particularly strong.

Our second  (and less significant) contribution applies to 
% a restricted class of 
polynomials with real coefficients: the roots of such polynomials
are either real or appear in complex conjugated pairs.
As a consequence, one can recover all the roots in a ROI $B_0$
containing $\R$ from the ones with positive imaginary parts.
We show how to improve a subdivision scheme by leveraging of the latter 
property.

Every polynomial $p$ and its product $p\conj{p}$ with its complex conjugate $\conj{p}$
belongs to this class and has additional property that
the multiplicity of its real roots is even, but we do not assume the latter 
restriction.

% \RI{
We implemented and tested our improvements in \ccluster.
For polynomials with real coefficients that are sparse or can 
be evaluated by a fast procedure, we achieved
a 2.5 to 3 fold speed-up as shown in table~\ref{table_intro} 
% }
% \RIn{
by columns $t_{old}/t_{new}$.
When the ROI contains only a few solutions,
\ccluster is, thanks to those improvements,
a little more efficient than \mpsolve
(compare columns \ccluster local, $t_{new}$
and \mpsolve in table~\ref{table_intro}).
We give details on our experiments below.
% }

\begin{table}[t!]
\centering
\begin{scriptsize}
 \begin{tabular}{l||c|ccc||c|ccc||c|}
                       & \multicolumn{4}{|c||}{\ccluster local}
                       & \multicolumn{4}{|c||}{\ccluster global}
                       & \multicolumn{1}{|c|}{\mpsolve } \\
                       & ~\#Clus~ & ~~$t_{old}$~ & ~$t_{new}$~~ & $t_{old}/t_{new}$
                       & ~\#Clus~ & ~~$t_{old}$~ & ~$t_{new}$~~ & $t_{old}/t_{new}$
                       & $t$\\\hline
% $\Mig{14}{64}$&1&0.01&0.00&2.24&63&0.73&0.31&2.34& 0.00\\\hline
$\Mig{14}{128}$~ &1&0.05&0.02&2.49&127&5.00&1.81&2.75& 0.02\\\hline
% $\Mig{14}{191}$&1&0.11&0.04&2.95&190&14.6&4.34&3.36& 0.04\\\hline
$\Mig{14}{256}$~ &1&0.16&0.05&2.82&255&31.8&10.7&2.95& 0.07\\\hline
$\Mig{14}{383}$~ &1&0.32&0.11&2.74&382&79.7&26.8&2.97& 0.17\\\hline\hline
% $\Man{6}$&0&0.00&0.00&1.67&63&0.99&0.44&2.20& 0.01\\\hline
$\Man{7}$~ &1&0.18&0.06&2.92&127&7.17&2.88&2.48& 0.06\\\hline
$\Man{8}$~ &0&0.39&0.11&3.38&255&40.6&15.1&2.69& 0.39\\\hline
$\Man{9}$~ &5&3.08&0.91&3.37&511&266&97.1&2.74& 3.20\\\hline
 \end{tabular}
 \caption{Running times in seconds of \ccluster, new and old versions,
          for computing clusters of roots in a small ROI (local)
          and  a ROI containing all the roots,
          and \mpsolve.
%          Initial box: $[-500, 500]+\ii [-500, 500]$.
%          $\epsilon=2^{-53}$.
         }
\label{table_intro}
\end{scriptsize}
\vspace*{-2em}
\end{table}

% \RI{
\vspace*{-1em}
\subsubsection*{Implementation and experiments.}

All the timings shown in this article
are sequential times in seconds on a
Intel(R) Core(TM) i7-7600U CPU @ 2.80GHz
machine with Linux.
\mpsolve is called with the command 
\texttt{mpsolve -as -Gi -o16 -j1}\footnote{
\mpsolve tries to isolate the roots unless the escape bound
$10^{-16}$ is reached.
}.
Table~\ref{table_intro} shows comparative running 
times of \ccluster and \mpsolve
on two families of polynomials, Mignotte
and Mandelbrot's polynomials, with real coefficients,
defined below.
Columns $t_{new}$ (resp. $t_{old}$) 
show timings of \ccluster with (resp. without)
the improvements described in this paper.
Columns \#Clus show the number of clusters found 
by two versions.
We used both versions of \ccluster with $\epsilon=2^{-53}$.
\ccluster global refers to the ROI $[-500, 500]+\ii [-500, 500]$,
that contains all the roots of the tested polynomials;
\ccluster local refers to an ROI containing only a few solutions.
We used $[-0.5, 0.5]+\ii [-0.5, 0.5]$
for Mignotte's polynomials and 
$[-0.25, 0.25]+\ii [-0.25, 0.25]$ for
Mandelbrot's polynomials.
% For \ccluster

\begin{figure}[t!]
     \vspace*{-1em}
	 \begin{minipage}{0.5\linewidth}
	 \centering
	  \includegraphics[width=6.5cm]{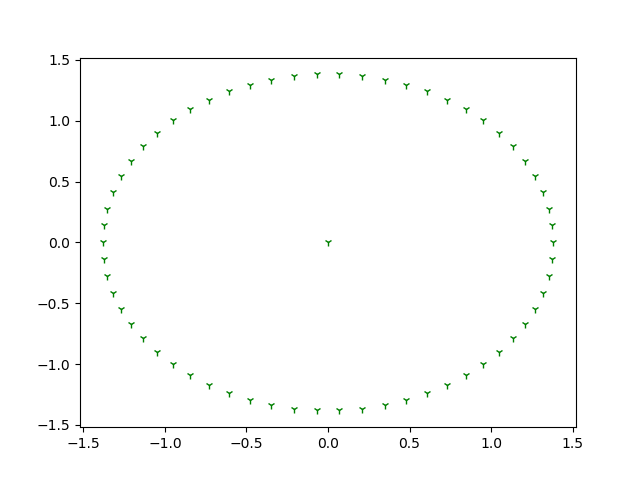}
	 \end{minipage}
	 \begin{minipage}{0.5\linewidth}
	 \centering
	  \includegraphics[width=6.5cm]{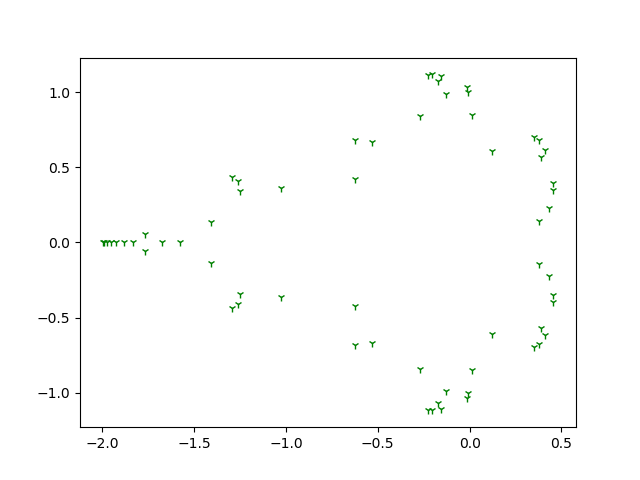} 
	 \end{minipage}
	    \caption{
	    {\bf Left:} 
	    $63$ clusters of roots for a Mignotte polynomial of degree 64.
	    {\bf Right:} Clusters of roots for the Mandelbrot polynomial of degree 63.
		  }
	 \label{fig:bernmand}
	\vspace*{-2em}
\end{figure}

The Mignotte's polynomial of degree $d$
and parameter $a=14$ is:

       \begin{equation}
       \label{eq:mignotte}
       \Mig{a}{d}(z) = z^d - 2(2^az-1)^2
       \end{equation}
It has a cluster of two roots near the origin
whose separation is near the theoretical minimum separation
bound. 
It is sparse and can be evaluated very fast.
% $63$ clusters of roots for $\Mig{14}{64}(z)$
% are drawn in Fig.~\ref{fig:bernmand}.
% It is sparse and can be evaluated very fast.
% 
We define the Mandelbrot's polynomial as $\Man{1}(z)=1$ and 

       \begin{equation}
       \label{eq:mandel}
       \Man{k}(z) = z\Man{k-1}(z)^2 + 1
       \end{equation}
% Then $\Man{d}(z)=p_{\lfloor log_2(d+1) \rfloor}(z)$.
$\Man{k}(z)$ has degree $2^{k}-1$.
It can be evaluated with
a straight line program.
The $63$ clusters of roots of $\Man{6}(z)$
and $\Mig{14}{64}(z)$
are depicted in Fig.~\ref{fig:bernmand}.

% }

\vspace*{-1em}
\subsubsection*{Structure of the paper.}

Our paper is organized as follows: in
Sec.~\ref{sec:counting} we describe our $\Pstart$-test.
\RIn{
In Sec.~\ref{sec:using} we apply it to speeding up a 
clustering algorithm.}
In Sec.~\ref{sec:real} we cover our root-finder for polynomials
with real coefficients.
Sec.~\ref{sec:results} presents the results of our improvements.
In the rest of the present section, 
we recall the related work and the 
clustering algorithm of \cite{2016Becker}.

\subsection{Previous works}

Univariate polynomial root-finding is a long-standing  
and still actual problem; it is intrinsically linked to 
numerical factorization of a polynomial into the product of its linear factors. The algorithms of \cite{pan2002univariate} support record and nearly
optimal bounds on the Boolean complexity of the solution of both problems of factorization and root-finding. 
The cost bound of the factorization is smaller by a factor of $d$, and both bounds 
  differ from respective information lower bound  by at most a polylogarithmic factor in the input size 
  and in the bound on the required output precision.
Root-finder supporting such bit
complexity bounds are said to be nearly optimal.
% \RIn{
The algorithms of \cite{pan2002univariate} are involved
and have never been implemented.
% }
User's choice
 has been for a while
the package of subroutines \mpsolve (see \cite{bini2000design} and \cite{bini2014solving}), based on simultaneous Newton-like 
(\emph{i.e.} Ehrlich-Aberth) iterations.
These iterations converge  to all roots simultaneously with cubic convergence rate, but only locally, that is, near the roots; empirically they converge very fast also globally, right from the start, although
 formal support for this empirical behavior
 is a long-known research challenge.
Furthermore these iterations compute a small number of roots in a ROI not much faster than all roots.

In contrast, recent approaches based on subdivision 
(as well as the algorithms of \cite{pan2002univariate})
compute the roots in a fixed ROI at a cost 
that decreases at least proportionally to the number of roots.
Near-optimal complexity has been achieved both for the 
real case (see \cite{pan2016nearly}, \cite{sagraloff2016computing}
that combines the Descartes rule of signs
with Newton's iterations and its implementation 
in \cite{Kobel})
and the complex case. 
In the complex case  \cite{becker+3:cisolate:18} similarly  
combines  counting test  based on Pellet's theorem
with complex version of the QIR algorithm, which in turn combines  Newton's and secant iterations.

\cite{2016Becker} extends the method of \cite{becker+3:cisolate:18}
for root clustering, \emph{i.e.} it solves the LCP
and  is robust
in the case of multiple roots; its implementation (\cite{ICMSpaper})
is a little more efficient than \mpsolve for ROI's
containing only several roots; when all the roots are sought,
\mpsolve remains the user's choice. 
The algorithms of \cite{2016Becker} and \cite{becker+3:cisolate:18} are direct successors of the previous subdivision algorithms of \cite{renegar1987worst} and \cite{pan2000approximating}, presented under the name of Quad-tree algorithms (inherited from the  earlier works by Henrici and Gargantini).

Besides Pellet's theorem, counting test in ROI
can rely on Eq.~(\ref{eq:cauchy}) and 
winding numbers algorithms (see, e.g.,
\cite{henrici1969uniformly,renegar1987worst} and  \cite{zaderman2019counting}).

\subsection{Solving the LCP}

% We present a simple version of the root clustering algorithm 
% of \cite{2016Becker}.
% \vspace*{-1em}
\subsubsection*{$\Czerot$ and $\Cstart$ tests}

%We consider
The two 
tests $\Czerot$ and $\Cstart$
discard boxes with no roots of $p$
and count the number of roots in a box,
respectively.
For a given complex disc $\Delta$,
% and a given polynomial $p$,
$\Czero{\Delta,p}$ returns either $-1$ or $0$, and returns $0$
only if $p$ has no root in $\Delta$,
while $\Cstar{\Delta,p}$ returns an integer $k\geq -1$
such that 
$k\geq 0$ only if $p$ has $k$ roots in $\Delta$.
%VP counted with multiplicity.
Below, we may write $\Czero{\Delta}$ for $\Czero{\Delta,p}$
and $\Cstar{\Delta}$ for $\Cstar{\Delta,p}$.

In \cite{2016Becker,becker+3:cisolate:18,ICMSpaper},
both $\Czerot$ and $\Cstart$ are based on the so called
``soft Pellet test'' denoted $\Tstar{\Delta,p}$
or $\Tstar{\Delta}$
which returns an integer $k\geq -1$ such that
$k\geq 0$ only if $p$ has $k$ roots in $\Delta$:
%VP counted with multiplicity:

\begin{equation}
\label{eq:C0C1TS}
\begin{array}{rl}
\Czero{\Delta} := &
\left\{ \begin{array}{rl}
         0  & \text{ if } \Tstar{\Delta} = 0 \\
         -1 & \text{ otherwise }\\
        \end{array}
\right. \\
 & \\
\Cstar{\Delta} := & \Tstar{\Delta}.
\end{array}
\end{equation}

\subsubsection*{Boxes, quadri-section and connected components}

The box $B$ centered in $c=a+\ii b$ with width $w$
is defined as $[a-w/2, a+w/2] + \ii [b-w/2, b+w/2]$.
We denote by $\width{B}$ the width of $B$.
We call \emph{containing disc} of $B$
    the disc $\contDisc{B}$ defined as $D(c,\frac{3}{4}\width{B})$.
We define the four children of $B$ as the four boxes centered in 
$(a\pm\frac{w}{4})+\ii (b\pm\frac{w}{4})$
with width $\frac{w}{2}$.

Recursive subdivisions of a ROI $B_0$ falls back
to the construction of a tree rooted in $B_0$.
Hereafter we refer to boxes that are nodes 
(and possibly leafs)
of this tree as the boxes of the subdivision tree of $B_0$.

A \emph{component} $\ccomp{C}$ is a set of connected boxes.
The component box $\compBox{\ccomp{C}}$ of a component 
$\ccomp{C}$ is a smallest square box
subject to $\ccomp{C}\subseteq\compBox{\ccomp{C}}\subseteq B_0$,
where $B_0$ is the initial ROI.
We write $\contDisc{\ccomp{C}}$ for $\contDisc{\compBox{\ccomp{C}}}$
and $\width{\ccomp{C}}$ for $\width{\compBox{\ccomp{C}}}$.
% The width $\width{\ccomp{C}}$ of a component $\ccomp{C}$ 
% is the width of $\compBox{\ccomp{C}}$.
Below we consider components made up of
boxes of the same width;
such a component is \emph{compact} if 
$\width{\ccomp{C}}$ is at most $3$ times the width of its
boxes. Finally, a component $\ccomp{C}$ is \emph{separated} 
from a set $S$ if 
$\forall \ccomp{C}'\in S, 4\contDisc{\ccomp{C}}\cap\ccomp{C}'=\emptyset$
and $4\contDisc{\ccomp{C}}\subseteq 2B_0$.

% \vspace*{-1em}
\subsubsection*{A root clustering algorithm}

We give in Algo.~\ref{algo:RCA} a simple
root clustering algorithm based on subdivision of ROI $B_0$.
For convenience we assume that
$p$ has no root in $2B_0\setminus B_0$
% where  $B_0$ denotes an initial ROI, 
but this limitation
can easily be removed.
The paper  \cite{2016Becker} proves that 
Algo.~\ref{algo:RCA} terminates and output correct 
solution provided that
% \RI{
the $\Czerot$ and $\Cstart$-tests are as in Eq.~(\ref{eq:C0C1TS}).
% }

% \RI{
Note that in 
the {\bf while} loop of Algo.~\ref{algo:RCA}, components
with widest containing box are processed first;
together with the definition of a separated component,
this implies the following remark:
\begin{Remark}
 \label{rem:separated}
 Let $\ccomp{C}$ be a component in Algo.~\ref{algo:RCA}
 that passes the test in step 4.
 Then $\ccomp{C}$ satisfies 
 $\nbSolsIn{\contDisc{\ccomp{C}}}{p}=
  \nbSolsIn{4\contDisc{\ccomp{C}}}{p}$.
\end{Remark}
% }

\begin{algorithm}[t!]
	\begin{algorithmic}[1]
	\caption{Root Clustering Algorithm}
	\label{algo:RCA}
	\Require{A polynomial $p\in\C[z]$, a ROI $B_0$, $\epsilon>0$;
	         suppose $p$ has no roots in $2B_0\setminus B_0$} 
	\Ensure{Set $R$ of components solving the LCP.
	}
% 	
% 	\coblue{{\it // Initialization}}
	\State $R\ass\emptyset$, $Q\ass\{B_0\}$ \coblue{{\it // Initialization}}
% 	
% 	\coblue{{\it // Main loop}}
	\While{$Q$ is not empty} \coblue{{\it // Main loop}}
        \State $\ccomp{C}\ass Q.pop()$ {\it //$\ccomp{C}$ has the widest containing box in $Q$}

        \coblue{{\it // Validation}}
        \If{ $\width{\ccomp{C}}\leq \epsilon$ {\bf and} 
             $\ccomp{C}$ is compact {\bf and} 
             $\ccomp{C}$ is separated from $Q$}
               \State $k\ass\Cstar{\contDisc{\ccomp{C}},p}$ 
               \If{ $k>0$ }
                   \State $R.push( (\ccomp{C},k) )$
                   \State {\bf break}
               \EndIf
        \EndIf
        
        \coblue{{\it // Bisection}}
        \State $S\ass$ empty set of boxes
        \For{ each box $B$ of $\ccomp{C}$}
        \For{ each child $B'$ of $B$}
                \If{$\Czero{\contDisc{B'},p}$ returns $-1$}
                    \State $S.push(B')$
                \EndIf
% %             \EndIf
        \EndFor
	\EndFor
	\State $Q.push($ connected components in $S$ $)$
    \EndWhile
	\State \Return $R$
	\end{algorithmic}
	\end{algorithm}
	
%%%%%%%%%%%%%%%%%%%%%%%%%%%%%%%%%%%%%%%%%%%%%%
\section{Counting the number of roots in a well isolated disk}
%%%%%%%%%%%%%%%%%%%%%%%%%%%%%%%%%%%%%%%%%%%%%%
\label{sec:counting}

In this section we present a new test for counting 
the number of roots with multiplicity of $p$ 
in a disc $\Delta$ provided that the roots in $\Delta$ are well
isolated from the other roots of $p$.
Let us first formalize this notion:

\begin{Definition}[Isolation ratio]
A complex disc $\Delta$ has isolation ratio $\rho$ for 
a polynomial $p$ if 
$\rho>1$ and
$\solsIn{\frac{1}{\rho}\Delta}{p}=\solsIn{\rho\Delta}{p}$.
% that is, no root of $p$ lie in the annulus centered in $\mcenter{\Delta}$
% with lower and upper radii $\frac{1}{\rho}\radius{\Delta}$ and $\rho\radius{\Delta}$.
\end{Definition}

Let $\solsIn{\Delta}{p}=\{\alpha_1, \ldots, \alpha_{d_\Delta}\}$
and let $m_i$ be the multiplicity of $\alpha_i$.
The $h$-th power sum of the roots in $\Delta$ is the complex number

\begin{equation}
 \label{eq:ps_def}
 s_h=\sum\limits_{i=1}^{d_\Delta} m_i\times\alpha_i^h
\end{equation}
% Notice that $s_0$ is the number of roots of $p$ in $\Delta$ counted
% with multiplicity.

In our test, called hereafter $\Pstart$-test, we approximate the $0$-th 
power sum $s_0$ of the roots of $p$ in $\Delta$ 
equal to the number of roots of $p$ in $\Delta$
(counted with multiplicity).
We obtain precise  $s_0$ from $s_0^*$ where
 $p$ and 
its derivative $p'$ are evaluated
on only a small number of points on the
contour of $\Delta$.
For instance, if $\Delta$ has isolation ratio $2$
and $p$ has degree 500,
our test amounts to evaluating $p$ and $p'$
on $q=11$ points; 
$s_0$ is recovered from these values
in $O(q)$ arithmetic operations.

\ignore{
If $p$ and its derivative can be evaluated at a low
computational cost,
e.g., if $p$ is sparse or $p$ is
 defined by a recurrence as the Mandelbrot
polynomial 
(see \cite{bini2000design}[Eq.~(16)]
 or Eq.~(\ref{eq:mandel}) above),
our $\Pstart$-test has additional benefits 
versus the $\Tstart$-test.
(Both tests have formal support under bounds on isolation of the boundary circle from the roots, the bound is milder  for $\Pstart$-test.)
{\bf REMI, please check this claim}
}
If $p$ and its derivative can be evaluated at a low
computational cost,
e.g. when $p$ is sparse or $p$
is defined by a recurrence as the Mandelbrot
polynomial
(see \cite{bini2000design}[Eq.~(16)]
 or Eq.~(\ref{eq:mandel}) above),
our $\Pstart$-test can be substantially 
cheaper to apply than the $\Tstart$-test
presented above.
Notice however that it requires the isolation ratio
of $\Delta$ (or at least a lower bound) to be known.

\subsection{Approximation of the 0-th power sum of the roots in a disk}

\cite{schonhage1982fundamental} and \cite{pan2018old}
give formulas for approximating the powers sums $s_h$
of the roots in the unit disk.
Here we 
% are interested in 
compute 
$s_0$ in any complex disk $\Delta=D(c,r).$ 
% centered in $c$ with radius $\Delta$.

For a positive integer $q$, define

\begin{equation}
\label{eq:ps}
 s_0^* = \frac{r}{q}\sum\limits_{g=0}^{q-1} \omega^{g} \frac{p'(c+r\omega^g)}{p(c+r\omega^g)}
\end{equation}
where $\omega=e^{\frac{2\pi\ii }{q}}$ denotes a primitive $q$-th root of unity.

The theorem below shows that 
the latter number approximates the $0$-th power sum with an error 
that can be made as tight as desired by increasing $q$,
providing that $\Delta$ has isolation ratio noticeably exceeding 1.

\begin{Theorem}
\label{cor:app_ps}
 Let $\Delta$ have isolation ratio $\rho$ for $p$,
 let $\theta=1/\rho$,
 let $s_0$ be the $0$-th power sum of the 
%  $d_{\Delta}$ 
 roots of $p$ in $\Delta$,
 and let $s_0^*$ be defined as in eq.~\ref{eq:ps}. Then
 \begin{enumerate}[(i)]
  \item $|s_0^*-s_0| \leq \dfrac{d\theta^{q}}{1-\theta^{q}}$.
  \item Fix $e>0$. If $q=\lceil \log_{\theta}(\frac{e}{d+e}) \rceil$
        then $|s_0^*-s_0| \leq e$.
 \end{enumerate}

\end{Theorem}

% \begin{proof}[of Thm.~\ref{cor:app_ps}.]
\noindent{\bf Proof of Thm.~\ref{cor:app_ps}}:
Let $p_{\Delta}(z)$ be the polynomial $p(c+rz)$.
Thus $p_{\Delta}'(z)=rp'(c+rz)$ and Eq.~(\ref{eq:ps})
rewrites
$s_0^* = \frac{1}{q}\sum\limits_{g=0}^{q-1} \omega^{g} \frac{p_{\Delta}'(\omega^g)}{p_{\Delta}(\omega^g)}$.
In addition, the unit disk $D(0,1)$
has isolation ratio $\rho$ for $p_{\Delta}$
and contains $s_0$ roots of $p_{\Delta}$.
%VP counted with multiplicity.
% in $\Delta_{0,1}$.
Then apply Thm.~14 in \cite{pan2018old}
to $p_{\Delta}(z)$ to obtain $(i)$. 
$(ii)$ is a direct consequence of $(i)$.

\qed
% \end{proof}

% \medskip

 For example, if $\Delta$ has isolation ratio $2$, $p$ has degree 500
 and one wants to approximate $s_0$ with an error less than $1/4$,
 it suffices to apply formula in Eq.~(\ref{eq:ps})
 for $q=11$, that is to evaluate $p$ and its derivative $p'$
 at $11$ points.
 
 \subsection{Black box for evaluating a polynomial $p$ on an oracle number}
 
 Our goal is to give an effective
 description of our $\Pstart$-test; to this end, let us introduce
 the notion of \emph{oracle numbers} that correspond to black boxes
 giving arbitrary precision approximations of any complex number.
 Such oracle numbers can be implemented through arbitrary precision
 interval arithmetic or ball arithmetic.
 Let $\intbox{\C}$ be the set of complex intervals.
 If $\intbox{a}\in\intbox{\C}$, then $w(\intbox{a})$
 is the maximum width of real and imaginary parts of $\intbox{a}$.
 
 For a number $a\in\C$, we call \emph{oracle} for $a$ a function
 $\calO_a:\N\rightarrow\intbox{\C}$
 such that $a\in\calO_a(L)$ and $\width{\calO_a(L)}\leq2^{-L}$
 for any $L$.
 Let $\calO_{\C}$ be the set of oracle numbers.
 
%  Let $p\in\C[z]$.
%  , with degree $d$.
%  An \emph{oracle} for $p$ consists in $d+1$ oracles for the coefficients
%  of $p$.
%  
 For a polynomial $p\in\C[z]$, we call
 \emph{evaluation oracle} for $p$
 a function
 $\calI_p: (\calO_{\C},\N)\rightarrow\intbox{\C}$,
 such that if $\calO_a$ is an oracle for $a$
 and $L\in\N$, then
 $p(a)\in\calI_p( \calO_a, L )$
 and 
 $\width{\calI_p( \calO_a, L )}\leq 2^{-L}$.
 
 We consider evaluation oracles $\calI_p$ and $\calI_{p'}$
 for $p$ and its derivative $p'$.
 If $p$ is given by $d+1$ oracles for its coefficients,
 one can easily construct $\calI_p$ and $\calI_{p'}$
 by using for instance Horner's rule.
 However for some polynomials defined by a procedure, 
 for instance the Mandelbrot polynomial 
 (see Eq.~(\ref{eq:mandel})),
 one can construct 
 fast evaluation oracles $\calI_p$ and $\calI_{p'}$
 from the procedurial definition.

\subsection{The $\Pstart$-test}

 \begin{algorithm}[t!]
	\begin{algorithmic}[1]
	\caption{$\Pstar{\calI_p, \calI_{p'}, \Delta, \rho}$ }
	\label{algo:counting}
	\Require{$\calI_p$, $\calI_{p'}$
	         evaluation oracles for $p$ and $p'$,
	         $\Delta=D(c,r)$, $\rho>1$.
	         $p$ has degree $d$.} 
	\Ensure{an integer in $\{0,\ldots,d\}$}
	
	\State $L\ass 53$, $w\ass1$, $e\ass 1/4$, $\theta\ass 1/\rho$
	\State $q\ass \lceil \log_{\theta}(\frac{e}{d+e}) \rceil$
	
	\While{$w\geq 1/2$}
        \State Compute interval $\intbox{s_0^*}$ as 
               $\frac{r}{q}\sum\limits_{g=0}^{q-1} 
               \calO_{\omega^g}(L) 
               \frac{\calI_{p'}(\calO_{c+r\omega^g},L)}
               {\calI_p(\calO_{c+r\omega^g},L)}$
        \State $w\ass w(\intbox{s_0^*})$
        \State $L\ass 2*L$
	\EndWhile
	\State $\intbox{s_0}\ass \intbox{s_0^*} + [-1/4,1/4]+\ii [-1/4,1/4]$
	\State \Return the unique integer in $\intbox{s_0}$
	\end{algorithmic}
\end{algorithm}

Algo.~\ref{algo:counting}
counts the number of roots of $p$ in a disk $\Delta=D(c,r)$
having isolation ratio at least $\rho$.
For such a disk, any positive integer $q$
and any integer $0\leq g < q$, one has
$p(c+r\omega^g)\neq 0$.
As a consequence, 
there exist an $L'$ s.t
$\forall L\geq L', \forall 0\leq g\leq q-1, 0\notin\calI_p(\calO_{c+r\omega^g},L)$
and the intervals $\intbox{s_0^*}$ computed in step 4
of Algo.~\ref{algo:counting} 
have strictly decreasing width as of $L\geq L'$.
This shows the termination of Algo.~\ref{algo:counting}.
Its correctness is stated in the following proposition:
\begin{Proposition}
\label{prop:ps_correct}
 Let $k$ be the result of the call 
 $\Pstar{\calI_p, \calI_{p'}, \Delta, \rho}$.
 If $\Delta$ has isolation ratio at least  $\rho$ for $p$,
 then $p$ has $k$ roots in $\Delta$
 counted with multiplicity.
\end{Proposition}

% \begin{proof}[of Prop.~\ref{prop:ps_correct}]
\noindent\textbf{Proof of Prop.~\ref{prop:ps_correct}.}
Once the {\bf while} loop in Algo.~\ref{algo:counting}
terminates, the interval $\intbox{s_0^*}$
contains $s_0^*$ 
% defined in Eq.~(\ref{eq:ps})
and $\width{\intbox{s_0^*}}<1/2$.
% and has width strictly less than $1/2$.
In addition, by virtue of statement $(ii)$ of Thm.~\ref{cor:app_ps},
one has $|s_0^*-s_0|\leq 1/4$,
thus $\intbox{s_0}$ defined in 
step 7 
% of Algo.~\ref{algo:counting} 
% has 
% width strictly less than $1$ and contains $s_0$;
satisfies: $\width{\intbox{s_0}}<1$ and $s_0\in\intbox{s_0}$.
Since $\intbox{s_0}$ contains at most one integer,
$s_0$
is the unique integer in $\intbox{s_0}$,
and is equal to the number of roots 
% of $p$
in $\Delta$.
% ^VP  counted with multiplicity.

\qed
\section{Using the $\Pstart$-test in a subdivision framework}
%%%%%%%%%%%%%%%%%%%%%%%%%%%%%%%%%%%%%%%%%%%%%%
\label{sec:using}

Let us discuss the use of the 
$\Pstart$-test as $\Czerot$ and $\Cstart$-tests
in order to speed up Algo.~\ref{algo:RCA}.
Table.~\ref{table_ps} covers runs of Algo.~\ref{algo:RCA}
on Mignotte and Mandelbrot's polynomials.
$t$ is the running time
when $\Czerot$ and $\Cstart$ tests are defined by 
Eq.~(\ref{eq:C0C1TS}).
Columns nb show the respective numbers
of $\Czerot$ and $\Cstart$-tests performed,
column $t_0$ and $t_0/t$ 
(resp. $t_*$ and $t_*/t$) show time
and ratio of times
spent in $\Czerot$ (resp. $\Cstart$) tests
when it is defined by Eq.~(\ref{eq:C0C1TS}).

One can 
readily use the $\Pstart$-test 
to implement the $\Cstart$-test by defining

\begin{equation}
\label{eq:CSPS}
\Cstar{\Delta} := \Pstar{\calI_p, \calI_{p'}, 2\Delta, 2}
\end{equation}
Following Rem.~\ref{rem:separated},
the $\Cstart$-test is called in Algo.~\ref{algo:RCA}
for components $\ccomp{C}$ satisfying
$\nbSolsIn{\contDisc{\ccomp{C}}}{p}=\nbSolsIn{4\contDisc{\ccomp{C}}}{p}$.
Hence $2\contDisc{\ccomp{C}}$ has isolation ratio $2$
and by virtue of Prop.~\ref{prop:ps_correct},
$\Cstar{\contDisc{\ccomp{C}}}$ returns $r\geq0$ only if 
$\contDisc{\ccomp{C}}$ contains $r$ roots.

However this would not imply much improvements in itself.
Column $t_*'$ in table.~\ref{table_ps}
shows the time that would be spent in $\Cstart$-tests
if it was defined by Eq.~(\ref{eq:CSPS}):
it is far less than $t_*$, but 
the ratio of time 
spent in $\Cstart$-tests (see column $t_*/t$)
is very small.
In contrast, about 90\% of the running time of
Algo.~\ref{algo:RCA} is spent in $\Czerot$-tests
(see column $t_0/t$).
% in table~\ref{table_ps}).
We propose to use 
a modified version of the 
$\Pstart$-test as a filter in the $\Czerot$-test
to decrease its running time.

\begin{table}[t!]
\centering
\begin{scriptsize}
 \begin{tabular}{l||c|cc|cccc||c|cc|c|}
                        & \multicolumn{7}{|c||}{$\Czerot$-tests}
                        & \multicolumn{4}{|c| }{$\Cstart$-tests} \\\hline
                        &  \multicolumn{1}{c}{ }  
                             & \multicolumn{2}{c}{$\Tstart$-tests}
                             & \multicolumn{4}{c||}{$\PstartApp$-tests}
                        &  \multicolumn{1}{c}{ }    
                             & \multicolumn{2}{c}{$\Tstart$-tests}
                             & $\Pstart$-tests \\
                        & nb & $t_0$ & $t_0/t$ (\%) & $t_0'$ & $n_{-1}$ & $n_{-2}$ & $n_{err}$ 
                        & nb & $t_*$ & $t_*/t$ (\%) & $t_*'$ \\\hline
%  $\Ber{128}$~         & ~4732~ & 5.50 & 86.9 & 1.38 & 269 & 0 & 10 & ~129~ & 0.12 & 1.89 & 0.04 \\
%  $\Ber{256}$~         & ~9980~ & 36.3 & 87.8 & 7.61 & 561 & 0 & 20 & ~257~ & 0.88 & 2.12 & 0.24 \\\hline
 
 $\Mig{14}{128}$~  & ~4508~ & 4.73 & 90.9 & 0.25 & 276 & 0 & 12 & ~128~ & 0.07 & 1.46 & 0.01 \\
 $\Mig{14}{256}$~  & ~8452~ & 27.8 & 91.2 & 0.60 & 544 & 0 & 20 & ~256~ & 0.58 & 1.92 & 0.02 \\\hline
 
 $\Man{7}$~        & ~4548~ & 6.34 & 88.1 & 0.28 & 168 & 0 & 28 & ~131~ & 0.11 & 1.51 & 0.01 \\
 $\Man{8}$~        & ~8892~ & 35.6 & 88.4 & 0.67 & 318 & 0 & 57 & ~256~ & 0.69 & 1.71 & 0.03 \\
 \end{tabular}
 \caption{Details on runs of Algo.~\ref{algo:RCA} on Mignotte and Mandelbrot's polynomials.}
\label{table_ps}
\end{scriptsize}
\end{table}

\subsection{An approximate $\Pstart$-test}

 \begin{algorithm}[t!]
	\begin{algorithmic}[1]
	\caption{$\PstarApp{\calI_p, \calI_{p'}, \Delta, \rho}$ }
	\label{algo:countingApp}
	\Require{$\calI_p$, $\calI_{p'}$
	         evaluation oracles for $p$ and $p'$,
	         $\Delta=D(c,r)$, $\rho>0$.
	         $p$ has degree $d$.} 
	\Ensure{an integer in $\{-2,-1,0,\ldots,d\}$}
	\end{algorithmic}
\end{algorithm}

The approximate version of the $\Pstart$-test
is aimed at being applied to a disk $\Delta=D(c,r)$
with unknown isolation ratio.
Unless $\Delta$ has isolation ratio $\rho>1$,
the very unlikely case where
for some $0\leq g<q$, $p(c+r\omega^g)=0$,
leads to a non-terminating call of 
$\Pstar{\calI_p, \calI_{p'}, \Delta, \rho}$.
% Algo.~\ref{algo:counting}.
Also,
$\intbox{s_0}$ computed in step 7 of Algo.~\ref{algo:counting}
       could contain no integer or 
       an integer that is not $s_0$.
We define the $\PstartApp$-test specified in Algo.~\ref{algo:countingApp}
by modifying Algo.~\ref{algo:counting} as follows:
\begin{enumerate}
 \item after step 3, if an 
       $\calI_p(\calO_{c+r\omega^g},L)$
       contains 0, the result -2 is returned;
 \item step 7 is replaced with: 
       $\intbox{s_0}\ass \intbox{s_0^*} + [-1/2,1/2]+\ii [-1/2,1/2]$,
 \item after step 7, unless $\intbox{s_0}$
       contains a unique 
%        positive 
       integer,
%        contains 0 or two integers,
       the result -1 is returned.
\end{enumerate}

Modification 1 ensures termination 
when $\Delta$ does not have isolation ratio $\rho>1$.
With modification 2, $\intbox{s_0}$
can have width greater than 1 and contain more than one integer.
With modification 3, the $\PstartApp$-test
can return -1 which means that no conclusion can be 
made.
If $\PstarApp{\calI_p, \calI_{p'}, \Delta, \rho}$
returns a positive integer, this result has still to be checked,
for instance, with the $\Tstart$-test.

In table~\ref{table_ps},
column $n_{-2}$ (resp. $n_{-1}$) shows the number of times
$\PstarApp{\calI_p, \calI_{p'}, \Delta, 2}$
returns $-2$ (resp. $-1$)
when applied in place of $\Tstar{\Delta}$ in 
the $\Czerot$-test.
Column $n_{err}$ shows the number of times the conclusion
of $\PstartApp$ was wrong, and 
$t_0'$ shows the total time spent in $\PstartApp$-tests.

\subsection{Using the $\Pstart$ and $\PstartApp$-test in a subdivision framework}

Our improvement of Algo.~\ref{algo:RCA} 
is based on the following heuristic remarks.
First, it is very unlikely that 
$\PstarApp{\calI_p, \calI_{p'}, \Delta, 2}$
returns -2 (see column $n_{-2}$ in table~\ref{table_ps}).
Second, when $\Pstar{\calI_p, \calI_{p'}, \Delta, 2}$
returns $k\geq 0$, it is very likely that $\Delta$ contains 
$k$ roots counted with multiplicity
(see column $n_{err}$ in table~\ref{table_ps}).
% Thus we use the $\Pstart$-test as a filter in the $\Czerot$-test.

We define the $\Czerot$-test as follows:

\begin{equation}
\label{eq:C0PS}
\Czero{\Delta} :=
\left\{ \begin{array}{rl}
         -1 & \text{ if } \PstarApp{\calI_p, \calI_{p'}, \Delta, 2}\notin\{-2, 0\}, \\
         -1 & \text{ if } \PstarApp{\calI_p, \calI_{p'}, \Delta, 2}\in\{-2, 0\} \text{ and } \Tstar{\Delta}\neq 0, \\
         0  & \text{ if } \PstarApp{\calI_p, \calI_{p'}, \Delta, 2}\in\{-2, 0\} \text{ and } \Tstar{\Delta}= 0. \\
        \end{array}
\right.
\end{equation}
If $\Czero{\Delta}$ is defined in Eq.~(\ref{eq:C0PS}), 
it returns $0$ only if $\Delta$ contains no root.
Thus Algo.~\ref{algo:RCA} with $\Czerot$ and $\Cstart$-tests 
defined by Eqs.~(\ref{eq:C0PS}) and~(\ref{eq:CSPS}) 
is correct.

Remark now that 
if a square complex box $B$ of width $w$ does not contain root
and is at a distance at least $\frac{3}{2}w$ from a root,
then $\contDisc{B}$ has isolation ratio $2$,
and  $\PstarApp{\calI_p, \calI_{p'}, \contDisc{B}, 2}$
returns $0$ or $-2$. As a consequence,
the termination of Algo.~\ref{algo:RCA} with $\Czerot$ and $\Cstart$-tests 
defined in Eqs.~(\ref{eq:C0PS}) and~(\ref{eq:CSPS}) 
amounts to the termination of 
Algo.~\ref{algo:RCA} with $\Czerot$ and $\Cstart$
defined in Eq.~(\ref{eq:C0C1TS}).
% }

%%%%%%%%%%%%%%%%%%%%%%%%%%%%%%%%%%%%%%%%%%%%%%
\section{Clustering roots of polynomials with real coefficients}
%%%%%%%%%%%%%%%%%%%%%%%%%%%%%%%%%%%%%%%%%%%%%%
\label{sec:real}
	
\begin{figure}[t!]
	\vspace*{-1em}
	 \begin{minipage}{0.5\linewidth}
	 \centering
	  \includegraphics[width=6.5cm]{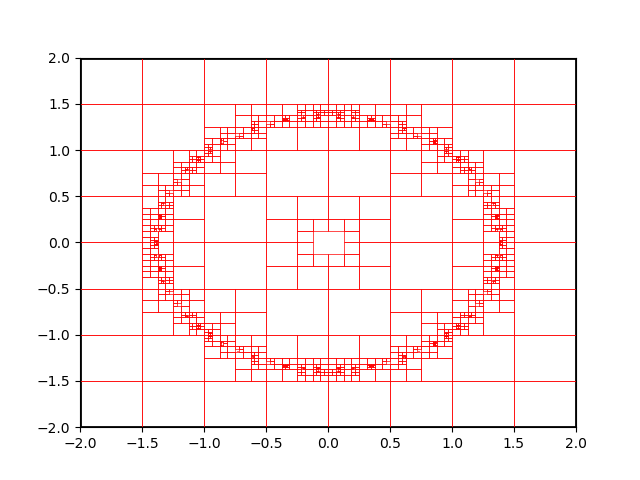}
	 \end{minipage}
	 \begin{minipage}{0.5\linewidth}
	 \centering
	  \includegraphics[width=6.5cm]{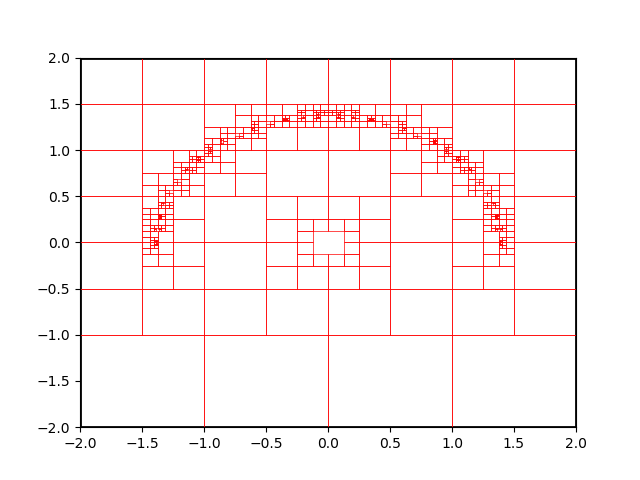} 
	 \end{minipage}
	    \caption{
	    Computing clusters for $\Mig{14}{64}$ in the ROI $[-2, 2]+\ii [-2, 2]$.
	    {\bf Left:} The subdivision tree for Algo.~\ref{algo:RCA}.
	    {\bf Right:} The subdivision tree for Algo.~\ref{algo:RCA_realcoeffs}.
		  }
	 \label{fig:mig_subdiv}
% 	\vspace*{-1em}
	\end{figure}

We consider here the special case where $p\in\R[z]$,
and show how to improve a subdivision 
algorithm for solving the LCP. 
We propose to
leverage on the geometric structure of the roots of $p$,
that are either real, or imaginary and come in
complex conjugated pairs:
if $\alpha\in\C$ is a root of $p$ so is $\conj{\alpha}$
where $\conj{\alpha}$ is the complex
conjugate of $\alpha$.
% Our improvements rely on this very basic property of complex 
% algebraic geometry.
The modified subdivision algorithm 
we propose deals only with
the boxes of the 
subdivision tree of the ROI $B_0$
that have a positive imaginary part;
the roots with positive imaginary parts
are in the latter boxes.
The roots with negative imaginary parts
are implicitly represented by the former ones.
In Fig.~\ref{fig:mig_subdiv} are shown two subdivision trees
constructed for clustering roots of a Mignotte polynomial 
of degree $64$; the left-most one is obtained when applying 
Algo.~\ref{algo:RCA}; the right-most one results of our improvement.

% For the sake of simplicity,
% we describe our algorithm in the cases where 
Below, we suppose that
$B_0$ is symmetric with respect
to the real axis
and that $p$ has no root in $2B_0\setminus B_0$.
These two limitations can easily be removed.

% If $B$ is a box centered in $c$, 
% we denote by $\conj{B}$, and call it the conjugate of $B$, the 
% box centered in $\conj{c}$ with width $\width{B}$.

% In what follow, we will consider subdivisions of
% an ROI $B_0$ that is symmetric with respect
% to the real axis:
% if $B$ is any box in the subdivision tree of $B_0$,
% its intersection with the real axis is either empty, 
% or equal to the lower or upper bounds of the imaginary 
% part of $B$.
% Moreover, $\conj{B}$ is also in the subdivision tree of $B_0$. 

\subsubsection*{Notations}

Let $B$ be a box centered in $c$.
We define its conjugate $\conj{B}$ as
the 
box centered in $\conj{c}$ with width $\width{B}$.
We say that $B$ is \emph{imaginary positive}
(resp. \emph{imaginary negative}) if $\forall b\in B$, $\mIm{b}>0$
(resp. $\mIm{b}<0$).

Let $\ccomp{C}$ be a component of 
boxes of the subdivision tree of $B_0$.
% , where $B_0$ is symmetric with respect to the real axis.
We define $\conj{\ccomp{C}}$ as the component which 
% constituent 
boxes
are the conjugate of the boxes of $\ccomp{C}$.
We call \emph{conjugate closure} of $\ccomp{C}$, 
and we denote it by $\conjclo{\ccomp{C}}$
the set of boxes $\ccomp{C}\cup(\conj{\ccomp{C}}\setminus \ccomp{C})$.
If $\ccomp{C}$ intersects $\R$, $\conjclo{\ccomp{C}}$ is a component.
We say that $\ccomp{C}$ is \emph{imaginary positive} (resp. \emph{imaginary negative}) 
if each box in $\ccomp{C}$ is 
imaginary positive (resp. imaginary negative).

% \subsection{Quadrisection of components}
% We present in 
% Algo.~\ref{algo:bisect_realcoeffs}
% our procedure to bisect a connected component.
% It discards boxes that are imaginary negative in addition to those 
% that contain no root.

\begin{algorithm}[t!]
	\begin{algorithmic}[1]
	\caption{$Quadrisect(\ccomp{C})$}
	\label{algo:bisect_realcoeffs}
	\Require{A polynomial $p\in\R[z]$ and a component $\ccomp{C}$} 
	\Ensure{A list $R$ of disjoint and not imaginary negative components}
	\State $S\ass$ empty list of boxes
	\For{ each constituent box $B$ of $\ccomp{C}$}
        \For{ each child $B'$ of $B$}
            \If{ $B$ is {\bf not} imaginary negative }
                \If{
%                  $\app{T}_0^G(\Delta(B'))$ fails 
%                  $\Tstar{\contDisc{B'}}\neq 0$
                 $\Czero{\contDisc{B'},p}$ returns -1
                 }
                    \State $S.push(B')$
                 \EndIf
            \EndIf
        \EndFor
	\EndFor
	\State $R\ass$ group boxes of $S$ in components
	\State \Return $R$
	\end{algorithmic}
	\end{algorithm}

\subsubsection*{Solving the LCP for polynomials with real coefficients}

\RI{
We describe in 
Algo.~\ref{algo:bisect_realcoeffs}
a procedure to bisect a component,
that discards boxes that are imaginary negative in addition to those 
that contain no root.
}

Our algorithm for solving the LCP for polynomials with real coefficients
is presented in Algo.~\ref{algo:RCA_realcoeffs}.
% As for Algo.~\ref{algo:RCA}, it is stated in the case where
% $2B_0\setminus B_0$ (where $B_0$ is the ROI)
% does not contain roots of $p$; this limitation
% can easily be removed.
It maintains in the queue $Q$ only components of boxes
% of the subdivision tree 
that are imaginary positive
or that intersect the real line.
Components with only imaginary negative boxes are 
implicitly represented by the imaginary positive ones.
Components that intersect the real line 
are replaced by their conjugate closure.
Components in $Q$ are ordered by decreasing width of their 
containing boxes.
% Roots with a negative imaginary part are implicitly 
% represented by these components.
% Components that intersect the real line 
% are replaced by their conjugate closure.
% When an imaginary positive component 
% is pushed in the result list $R$, its conjugate
% is also pushed in $R$.
The termination of Algo.~\ref{algo:RCA_realcoeffs}
is a consequence of the termination of Algo.~\ref{algo:RCA}
that is proved in \cite{2016Becker}.

% The following proposition emphasizes its correction.
Let $\{(\ccomp{C}^1,m^1),\ldots,(\ccomp{C},m^\ell)\}$ be the list returned by 
 Algo.~\ref{algo:RCA_realcoeffs} called for arguments $p,B_0,\epsilon$.
Then
$\{(\contDisc{\ccomp{C}^1},m^1),\ldots,(\contDisc{\ccomp{C}^\ell},m^\ell)\}$
 is a solution of the LCP problem for $p,B_0,\epsilon$, \emph{i.e.}:
\begin{enumerate}[$(i)$]
 \item the $\contDisc{\ccomp{C}^i}$'s are pairwise disjoint with radius less that $\epsilon$,
 \item $\forall 1\leq i \leq \ell$, $(\ccomp{C}^i,m^i)$ satisfies
$\nbSolsIn{\contDisc{\ccomp{C}^i}}{p}=\nbSolsIn{3\contDisc{\ccomp{C}^i}}{p}=m^i$,
 \item $\solsIn{B_0,p}{p} \subseteq
		\bigcup_{i=1}^{\ell} \solsIn{\contDisc{\ccomp{C}^i},p}{p} \subseteq
		\solsIn{2B_0,p}{p}$.
\end{enumerate}

In what follow we may write $R$ for the list of connected components in $R$.
$(i),(ii)$ and $(iii)$ are direct consequences of the following proposition:
\begin{Proposition}
 \label{prop_rc_correctness}
 Consider $Q$ and $R$ after any execution of the {\bf while} loop in Algo.~\ref{algo:RCA_realcoeffs}.
 Decompose $Q$ in two lists $Q^1$ and $Q^2$ containing respectively
 the imaginary positive components of $Q$ and the non imaginary components of $Q$.
 Note $\conj{Q^1}$ the list of the conjugates of the components in $Q^1$
 and $\conjclo{Q^2}$ the list of the conjugate closures of the components in $Q^2$,
 and let $\conjclo{Q}$ be $\conj{Q^1}\cup\conjclo{Q^2}$.
 One has:
 \begin{enumerate}[(1)]
  \item any $\alpha\in\solsIn{B_0}{p}$ is in $R\cup Q \cup\conjclo{Q}$,
  \item any $\ccomp{C}\in R$ is separated from $(R\setminus\{\ccomp{C}\})\cup Q \cup\conjclo{Q}$,
  \item any $(\ccomp{C},m)$ in $R$ is such that $m=\nbSolsIn{\contDisc{\ccomp{C}}}{p}=\nbSolsIn{3\contDisc{\ccomp{C}}}{p}$.
 \end{enumerate}

\end{Proposition}

% \RIn{
Proposition~\ref{prop_rc_correctness} is a consequence of 
Rem.~\ref{rem:separated} and the following remark.
% }
\begin{Remark}
 \label{rem_symetry}
 Let $p\in\R[z]$ and $\ccomp{C}$ be a component.
 If $\ccomp{C}$ is imaginary negative or imaginary positive and if
 there exists $m$ such that $m=\nbSolsIn{\contDisc{\ccomp{C}}}{P}=\nbSolsIn{3\contDisc{\ccomp{C}}}{P}$,
 then $m=\nbSolsIn{\contDisc{\conj{\ccomp{C}}}}{P}=\nbSolsIn{3\contDisc{\conj{\ccomp{C}}}}{P}$.
\end{Remark}

	\begin{algorithm}[!t]
	\begin{algorithmic}[1]
	\caption{Local root clustering for polynomials with real coefficients}
	\label{algo:RCA_realcoeffs}
	\Require{A polynomial $p\in\R[z]$, a ROI $B_0$, $\epsilon>0$;
	         assume $p$ has no roots in $2B_0\setminus B_0$,
	         and $B_0$ is symmetric with respect to the real axis.} 
	\Ensure{A set $R$ of components solving the LCP.
	}
	
% 	\coblue{{\it // Initialization}}
	\State $R\ass\emptyset$, $Q\ass\{\{B_0\}\}$ \coblue{{\it // Initialization}}
	
% 	\coblue{{\it // Main loop}}
	\While{$Q$ is not empty} \coblue{{\it // Main loop}}
        \State $\ccomp{C}\ass Q.pop()$ \coblue{\it //$\ccomp{C}$ has the widest containing box in $Q$}
        \State $sFlag\ass$ {\bf false}
        \If{$\ccomp{C}$ is {\bf not} imaginary positive } \coblue{\it //Note: $\ccomp{C}\cap\R\neq\emptyset$ }
            \State $\ccomp{C}\ass \conjclo{\ccomp{C}}$
            \State $sFlag\ass$ $\ccomp{C}$ is separated from $Q$
        \Else
            \State $sFlag\ass$ ($\ccomp{C}$ is separated from $Q$) 
                  {\bf and} ($4\contDisc{\ccomp{C}}\cap\conj{\ccomp{C}}=\emptyset$)
        \EndIf
%         
%         \coblue{{\it // Validation}}
        \If{ $\width{\ccomp{C}}\leq \epsilon$ {\bf and} 
             $\ccomp{C}$ is compact {\bf and} 
             $sFlag$} \coblue{{\it // Validation}}
               \State $m\ass \Cstar{\contDisc{\ccomp{C}},p}$
               \If{ $m>0$ }
                   \State $R.push( (\ccomp{C},m) )$
                   \If{$C$ is imaginary positive}
                        \State $R.push( (\conj{\ccomp{C}},m) )$
                   \EndIf
                   \State {\bf break}
               \EndIf
        \EndIf
%         
%         \coblue{{\it // Bisection}}
        \State $Q.push(Quadrisect(\ccomp{C}))$ \coblue{{\it // Bisection}}
	\EndWhile
	\State \Return $R$
	\end{algorithmic}
	\end{algorithm}

%%%%%%%%%%%%%%%%%%%%%%%%%%%%%%%%%%%%%%%%%%%%%%
\section{Numerical results}
%%%%%%%%%%%%%%%%%%%%%%%%%%%%%%%%%%%%%%%%%%%%%%
\label{sec:results}

% \RI{
We implemented the two improvements
of Secs.~\ref{sec:using} and~\ref{sec:real}
in \ccluster.
\cclusterO refers to the original version of \ccluster.
Both \cclusterR and \cclusterPs
implement Algo.~\ref{algo:RCA_realcoeffs}.
In \cclusterPs,
$\Czerot$ and $\Cstart$ are defined by 
Eqs.~(\ref{eq:C0PS}) and~(\ref{eq:CSPS}).
% The three versions of \ccluster have been called with
% ROI $B_0=[-500, 500]+\ii [-500, 500]$
% and $\epsilon=2^{-53}$.

\vspace*{-1em}
\subsubsection*{Testing suite.}
\begin{figure}[t!]
	\vspace*{-1em}
	 \begin{minipage}{0.5\linewidth}
	 \centering
	  \includegraphics[width=6.5cm]{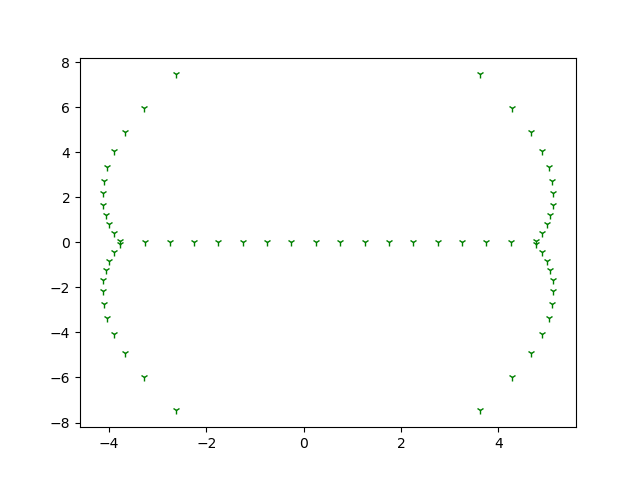}
	 \end{minipage}
	 \begin{minipage}{0.5\linewidth}
	 \centering
	  \includegraphics[width=6.5cm]{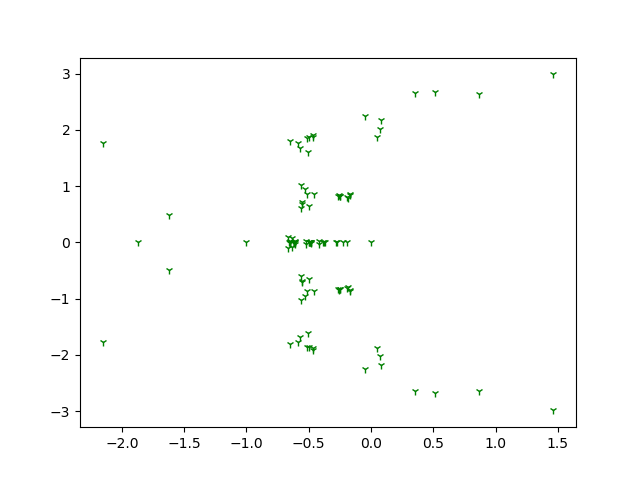} 
	 \end{minipage}
	    \caption{
	    {\bf Left:} 
	    64 clusters of roots for the Bernoulli polynomial of degree 64.
	    {\bf Right:} 107 clusters of roots for the Runnel's polynomial of degree 170.
		  }
	 \label{fig:migrun}
	\vspace*{-1em}
	\end{figure}

We tested our improvements
on Mignotte and Mandelbrot's polynomials
and on Bernoulli and Runnel's polynomials:
the Bernoulli polynomial of degree $d$
is $\Ber{d}(z)=\sum_{k=0}^{d} {{d}\choose{k}}b_{d-k}z^k$
where the $b_i$'s are the Bernoulli numbers.
It has about $d/2$ non-zero coefficients and, 
as far as we know,
cannot be evaluated substantially faster than
with Horner's scheme.
It has real coefficients, and about $2/3$ of its roots
are real or imaginary positive
(see left part of Fig.~\ref{fig:migrun}).
Let $r=2$, $q_0(z)=1$, $q_1(z)=z$ and
$q_{k+1}(z) = q_k(z)^r + zq_{k-1}(z)^{r^2}$.
We define the Runnel's polynomial of parameter $k$
as $\Run{k}=q_k$.
It has real coefficients, a multiple root (zero),
and can be evaluated fast.
The $107$ distinct roots of $\Run{8}$
are drawn on right part of Fig.~\ref{fig:migrun}.
	
\vspace*{-1em}
\subsubsection*{Results.}
Table.~\ref{table_real} gives details concerning the execution 
of \cclusterO, \cclusterR and \cclusterPs
for polynomials with increasing 
degrees.
We used $\epsilon=2^{-53}$ and the ROI 
$B_0=[-500, 500]+\ii [-500, 500]$ that 
contains all the roots of all the considered polynomials.
Column (\#Clus,\#Sols) shows
the number of clusters and the total multiplicity
found.
Columns (depth, size) show the depth and the 
size (\emph{i.e.} number of nodes) of 
the subdivision tree for each version.
$t_1$, $t_2$ and $t_3$ stand respectively 
for the running time in second of \cclusterO,
\cclusterR and \cclusterPs.

   Algo.~\ref{algo:RCA_realcoeffs} achieves   
    speed up $t_1/t_2$.
It is almost 2 for Mignotte polynomials, since about 
half of
its roots
are above the real axis. 
This speed up is less
important for the three other families of polynomials,
which have a non-negligible ratio of real roots.
% % % % % 
The speed up achieved by using the $\Pstart$-test is $t_2/t_3$.
It is significant for Mignotte's polynomial, which is sparse, 
and Mandelbrot and Runnel's polynomials
for which one can construct fast evaluation procedures.

\begin{table}[t]
% \hspace{-2cm}
\begin{scriptsize}
\begin{tabular}{l||c|c|c||c|c||c|c|c|c||}
                         & \multicolumn{3}{|c||}{\cclusterO}
                         & \multicolumn{2}{|c||}{\cclusterR}
                         & \multicolumn{4}{|c||}{\cclusterPs}\\
               & (\#Clus, \#Sols) & (depth, size) & $t_1$ & (depth, size) & $t_1/t_2$  & (depth, size) & $t_3$ & $t_2/t_3$  & $t_1/t_3$    \\\hline
% $\Ber{64}$     & (64, 64)         & (90, 2492)    & 0.99  & (90, 2004)    & 1.64   & (90, 2132)    & 0.62   & .964   &1.58      \\\hline
$\Ber{128}$    & (128, 128)       & (100, 4732)   & 6.30  & (100, 3708)   & 1.72   & (100, 4104)   & 3.30   & 1.10   &1.90      \\\hline
$\Ber{191}$    & (191, 191)       & (92, 7220)    & 20.2  & (92, 5636)    & 1.74   & (92, 6236)    & 10.7   & 1.08   &1.88      \\\hline
$\Ber{256}$    & (256, 256)       & (93, 9980)    & 41.8  & (93, 7520)    & 1.67   & (91, 8128)    & 21.9   & 1.14   &1.90      \\\hline
$\Ber{383}$    & (383, 383)       & (93, 14504)   & 120   & (93, 11136)   & 1.82   & (93, 11764)   & 53.5   & 1.23   &2.25      \\\hline\hline
% $\Mig{14}{64}$ & (63, 64)         & (92, 2044)    & 0.73  & (92, 1432)    & 1.85   & (94, 1516)    & 0.31   & 1.26   &2.34      \\\hline
$\Mig{14}{128}$& (127, 128)       & (96, 4508)    & 5.00  & (92, 3212)    & 1.92   & (92, 3484)    & 1.81   & 1.43   &2.75      \\\hline
$\Mig{14}{191}$& (190, 191)       & (97, 6260)    & 15.5  & (97, 4296)    & 2.01   & (97, 4688)    & 4.34   & 1.77   &3.58      \\\hline
$\Mig{14}{256}$& (255, 256)       & (94, 8452)    & 31.8  & (94, 5484)    & 2.04   & (94, 6648)    & 10.7   & 1.44   &2.95      \\\hline
$\Mig{14}{383}$& (382, 383)       & (97, 12564)   & 79.7  & (97, 8352)    & 1.98   & (97, 9100)    & 26.8   & 1.49   &2.97      \\\hline\hline
% $\Man{6}$      & (63, 63)         & (93, 2200)    & 0.99  & (93, 1500)    & 1.69   & (92, 1720)    & 0.44   & 1.30   &2.20      \\\hline
$\Man{7}$      & (127, 127)       & (96, 4548)    & 7.17  & (96, 2996)    & 1.62   & (96, 3200)    & 2.88   & 1.52   &2.48      \\\hline
$\Man{8}$      & (255, 255)       & (96, 8892)    & 40.6  & (96, 5576)    & 1.71   & (96, 6208)    & 15.1   & 1.56   &2.69      \\\hline
$\Man{9}$      & (511, 511)       & (100, 17956)  & 266   & (100, 11016)  & 1.89   & (100, 11868)  & 97.1   & 1.44   &2.74      \\\hline\hline
% $\Run{7}$      & (54, 85)         & (94, 2460)    & 2.15  & (94, 1804)    & 1.58   & (94, 1884)    & 0.97   & 1.39   &2.20      \\\hline
$\Run{8}$      & (107, 170)       & (96, 4652)    & 13.3  & (96, 3252)    & 1.61   & (96, 3624)    & 6.51   & 1.26   &2.04      \\\hline
$\Run{9}$      & (214, 341)       & (99, 9592)    & 76.2  & (99, 6260)    & 1.70   & (99, 6624)    & 32.2   & 1.38   &2.36      \\\hline
$\Run{10}$     & (427, 682)       & (100, 19084)  & 479   & (100, 12288)  & 1.69   & (100, 12904)  & 211    & 1.34   &2.26      \\\hline
\end{tabular}
\caption{Details on runs of \cclusterO,  \cclusterR
         and \cclusterPs for polynomials in $\R[z]$ 
         with increasing degree.
%          $d$.
% %          Initial box: $[-500, 500]+\ii [-500, 500]$.
%          $\epsilon=2^{-53}$.
         }
\end{scriptsize}
\label{table_real}
\end{table}
% }
% \subsection{Root clustering for polynomial known as black box}

%%%%%%%%%%%%%%%%%%%%%%%%%%%%%%%%%%%%%%%%%%%%%%
\section{Future works}
%%%%%%%%%%%%%%%%%%%%%%%%%%%%%%%%%%%%%%%%%%%%%%
\RIn{
Our main contribution  is a significant practical progress in
subdivision root-finding based on 
a new test for counting roots
in a 
 well-isolated disc.
If the latter assumption does not hold,
the test result is not guaranteed 
but is very likely to be correct.
In a subdivision framework,
we have proposed to use a test based on
Pellet's theorem to verify its result.
We aim to do so
by using only evaluations 
of $p$ and $p'$.
This would imply a very significant improvement
of the root clustering algorithm
 when $p$ and $p'$
can be evaluated very efficiently.
}

%%%%%%%%%%%%%%%%%%%%%%%%%%%%%%%%%%%%%%%%%%%%%%
\bibliographystyle{alpha}
%%%%%%%%%%%%%%%%%%%%%%%%%%%%%%%%%%%%%%%%%%%%%%
% \bibliography{../Latex_stuff/bib-algcurves}
% \bibliographystyle{alpha}
\bibliography{references}

%%%%%%%%%%%%%%%%%%%%%%%%%%%%%%%%%%%%%%%%%%%%%%
\end{document}